\documentclass[pre,preprint,showpacs,eqsecnum]{revtex4}
\usepackage{graphicx}
\begin{document}
\title{Arrival time distribution for a driven system containing quenched
dichotomous disorder}
\author{S.~I.~Denisov,$^{1,2}$ M.~Kostur,$^{1}$ E.~S.~Denisova,$^{2}$
and P.~H\"{a}nggi$^{1}$} \affiliation{$^{1}$Institut f\"{u}r Physik,
Universit\"{a}t Augsburg, Universit\"{a}tsstra{\ss}e 1, D-86135 Augsburg,
Germany\\
$^{2}$Sumy State University, 2 Rimsky-Korsakov Street, 40007 Sumy, Ukraine}

%\date{submitted to Physical Review E: \today}

\begin{abstract}
We study the arrival time distribution of overdamped particles driven by a
constant force in a piecewise linear random potential which generates the
dichotomous random force. Our approach is based on the path integral
representation of the probability density of the arrival time. We explicitly
calculate the path integral for a special case of dichotomous disorder and use
the corresponding characteristic function to derive  prominent properties of
the arrival time probability density. Specifically, we establish the scaling
properties of the central moments, analyze the behavior of the probability
density for short, long, and intermediate distances. In order to quantify the
deviation of the arrival time distribution from a Gaussian shape, we evaluate
the skewness and the kurtosis.
\end{abstract}

\maketitle

\section{Introduction}

The overdamped equation of motion of a classical particle in one dimension
presents a simple yet very useful model for the study of many physical,
biological, economical, and other systems. Depending on the character of the
force field acting on the particle, this equation provides a basis for
describing different phenomena in these systems. Specifically, if the force
field contains the noise terms arising from the influence of the environment,
it describes a large variety of noise-induced phenomena including noise-induced
transitions \cite{HL}, directed transport \cite{Trans}, and stochastic
resonance \cite{GHJM}, to name only a few. It should be noted that in some
cases, especially within the white-noise approximation, the statistical
properties of the solution of this equation can be obtained analytically.

If the environment is disordered the force field contains also the random
functions of the spatial variable. In this case the overdamped dynamics
represents both the noise-induced and disorder-induced effects and it can
exhibit as well anomalous behavior even in the simplest situation of additive
white noise \cite{BG}. For the latter situation a number of exact results was
obtained for Sinai disorder \cite{BG, Sin, Der, Gol, Mon}, Gaussian disorder
\cite{HTB, Sch, DV, GB, LV}, and also for some special cases of non-Gaussian
disorder \cite{Den, DL, PKDK, KKDP, DH}.

When the noise terms produced by a stochastic environment become negligible,
the overdamped equation of motion accounts solely for effects of quenched
disorder. This equation effectively describes, e.g., the transport of particles
in deterministic ratchets with quenched disorder \cite{PASF, GLZH, ZLAF} and
can be used for the study of the dynamics of localized structures like domain
walls in random magnets and vortices in type-II superconductors. Although
temporal noise terms are absent, there are only very few exact results
available. Therefore, in order to fill this gap, we have examined the following
dimensionless equation of motion for an overdamped particle \cite{DKDH}:
\begin{equation}
    \dot{X}_{t} = f + g(X_{t}).
    \label{eq motion}
\end{equation}
Here, $X_{t}$ denotes the particle coordinate that satisfies the initial
condition $X_{0} = 0$, $f\, (>0)$ is a constant force, and $g(x) = -dU(x) / dx
= \pm g$ is a dichotomous random force generated by a piecewise linear random
potential $U(x)$, see Fig.~1. It is assumed that the random intervals $s_{j}$
of a linearly varying $U(x)$ are statistically independent and distributed with
the same (exponential) probability density $p(s)$. Moreover, we assume that the
conditions $f> g$ and $g(+0) = -g$ are imposed.

Equation (\ref{eq motion}) is of minimal form that accounts for the effects of
quenched disorder on the overdamped motion of driven particles. Its main
advantage is that many of the statistical properties of $X_{t}$ can be
described analytically in full detail. Nevertheless, if the odd and even
intervals $s_{j}$ are distributed with different exponential densities (in this
case the exact results exist as well), then Eq.~(1.1) can be used also for
studying a number of important physical issues. Specifically, this equation
constitutes a basis for describing the adiabatic transport of particles in
randomly perturbed one-dimensional channels and presents a simple model for
studying the low-temperature dynamics of charge carriers and localized
structures in randomly layered media. In addition to the listed examples, we
point out at rather unexpected application of Eq.~(1.1) in astrophysics.
Namely, if the clouds in interstellar space are distributed uniformly then the
distances between them are distributed with an exponential distribution. In
this case, assuming that the light velocity in the clouds is the same and the
sizes of clouds are distributed with an exponential distribution, Eq.~(1.1) can
be used for studying the statistical properties of distances that pass the
light emitted by a star in different directions.

In \cite{DKDH} we derived the probability density of the solution of
Eq.~(\ref{eq motion}) and investigated explicitly its time evolution. In
contrast, in this work we focus on the statistical properties of the arrival
time for the particles governed by Eq.~(\ref{eq motion}). The paper is
structured as follows. In Sec.~II, we derive the path integral representation
for the probability density of the arrival time. The characteristic function of
the arrival time is determined in Sec.~III. In Sec.~IV, we calculate the
moments of the arrival time and study their asymptotic and scaling behavior.
The basic properties of the arrival time probability density are studied in
Sec.~V, both analytically and numerically. We summarize our findings in Sec~VI.
Some technical details of our calculations are deferred to the Appendix.

\section{PATH INTEGRAL REPRESENTATION OF THE ARRIVAL TIME PROBABILITY DENSITY}

According to Eq.~(\ref{eq motion}), the arrival time $t_{x}$, i.e., the time
that a particle spends moving from the origin to a position $x(>0)$, is given
by the integral expression
\begin{equation}
    t_{x} = \int_{0}^{x}\frac{dx}{f + g(x)}.
    \label{arr time}
\end{equation}
This time depends on the random function $g(x)$ and thus presents a random
quantity. The probability density $P_{x}(t)$ that $t_{x} = t$ for a fixed
coordinate $x$, i.e., the probability density of the arrival time, is defined
in the well-known way as
\begin{equation}
    P_{x}(t) = \langle \delta(t - t_{x}) \rangle,
    \label{def P(t)}
\end{equation}
where the angular brackets denote an averaging over the sample paths of $g(x)$,
and $\delta(t - t_{x})$ is the Dirac $\delta$ function.

To obtain  the explicit form of $P_{x}(t)$ we use a path integral approach.
Because of the dichotomous character of the random function $g(x)$, it is
convenient to present the probability density in terms of the partial densities
\begin{equation}
    P_{x}(t) = \sum_{n=0}^{\infty}P_{x}^{(n)}(t),
    \label{Px(t)1}
\end{equation}
where $P_{x}^{(n)}(t)$ is the probability density that for the sample paths of
$g(x)$ which are undergoing $n$ changes of the sign on the interval $(0,x)$ the
condition $t_{x} = t$ holds. For a given $n\,(\geq 1)$ the solution of
Eq.~(\ref{eq motion}) can be written in the form
\begin{equation}
    X_{t}^{(n)} = \sum_{j = 1}^{n}s_{j} + \tilde{s}_{n+1}
    \label{Xt}
\end{equation}
with $\tilde{s}_{n+1} \in (0,s_{n+1})$. On the other hand, because $g(x) =
(-1)^{j}g$ if $x$ belongs to the interval $s_{j}$, (\ref{arr time}) yields
\begin{equation}
    t_{x}^{(n)} = \sum_{j=1}^{n}\frac{s_{j}}{f + (-1)^{j}g} +
    \frac{\tilde{s}_{n+1}} {f + (-1)^{n+1}g}.
    \label{rel1}
\end{equation}
Setting $X_{t}^{(n)} = x$ and replacing $\tilde{s}_{n+1}$ by $x - \sum_{j=1}
^{n}s_{j}$, the result (\ref{rel1}) can be recast to
\begin{equation}
    t_{x}^{(n)} = \frac{x}{f - (-1)^{n}g} - \frac{g}{f^{2} - g^{2}}
    \sum_{j=1}^{n}[(-1)^{n} + (-1)^{j}]s_{j}.
    \label{rel1a}
\end{equation}

Let us next introduce the probability $p(s_{j})ds_{j}$ that the $j$-th jump of
$g(x)$ occurs in the interval $ds_{j}$ and also the probability $\int_{l}^
{\infty} p(s)ds$ that the distance between the nearest-neighbor jumps exceeds
$l$. Then, the probability $dW_{n}(x)$ that the function $g(x)$ on the interval
$(0,x)$ experiences $n$ jumps in the intervals $ds_{j}$ ($j = 1,\ldots,n$)
assumes the form
\begin{equation}
    dW_{n}(x) = \int_{x - \sum_{j=1}^{n}s_{j}}^{\infty}p(s)ds
    \prod_{j=1}^{n}p(s_{j})ds_{j}.
    \label{dWn}
\end{equation}
Because $\tilde{s}_{n+1} > 0$, the positive variables of integration, $s_{j}$,
must satisfy the condition $\sum_{j=1}^{n}s_{j} < x$. Denoting by $\Omega_{n}
(x)$ a region in the $n$-dimensional space of these variables, being defined by
the aforementioned condition, we obtain
\begin{equation}
    P_{x}^{(n)}(t) = \int_{\Omega_{n}(x)}\delta(t - t_{x}^{(n)})
    dW_{n}(x).
    \label{P^n(t)}
\end{equation}
Finally, taking into account that $t_{x}^{(0)} = x/(f-g)$ is the arrival time
at $n = 0$ and $W_{0}(x) = \int_{x} ^{\infty}p(s)ds$ is the total probability
of those sample paths of $g(x)$ which do not change  sign on the interval
$(0,x)$, we end up with the following path integral representation for the
probability density of the arrival time:
\begin{equation}
    P_{x}(t) = \delta(t - t_{x}^{(0)})W_{0}(x) + \sum_{n=1}^{\infty}
    \int_{\Omega_{n}(x)}\delta(t - t_{x}^{(n)})dW_{n}(x).
    \label{Px(t)2}
\end{equation}

This form of the arrival time probability density is rather general, but
possesses  a rather complex mathematical structure. Based on (\ref{Px(t)2}),
however, we arrive at two conclusions that are valid for an arbitrary
probability density $p(s)$: (i) $P_{x}(t)$ at a fixed $x$ is concentrated on
the interval $[x/(f+g), x/(f-g)]$, and (ii) $P_{x}(t)$ is properly normalized,
i.e., $\int_{0}^ {\infty} P_{x}(t)dt = 1$. Indeed, since $\min{t_{x}^{(n)}} =
x/(f+g)$ and $\max{t_{x}^{(n)}} = x/(f-g)$, we have $\delta(t - t_{x}^{(n)})
\equiv 0$ and so $P_{x}(t) \equiv 0$ if $t \notin [x/(f+g), x/(f-g)]$. To prove
the second assertion, we first note that, according to (\ref{Px(t)2}),
$\int_{0}^ {\infty} P_{x}(t)dt = W_{0}(x) + \sum_{n=1}^{\infty} W_{n}(x)$,
where $W_{n}(x) = \int_{\Omega_{n} (x)} dW_{n}(x)$ is the probability that the
function $g(x)$ has undergone $n$ jumps in the interval $(0,x)$. Next,
introducing the quantities $S_{n}(x) = \int_{\Omega_{n}(x)} \prod_{j=1}^{n}
p(s_{j})ds_{j}$, we find the representations $W_{0}(x) = 1 - S_{1}(x)$ and
$W_{n}(x) = S_{n}(x) - S_{n+1} (x)$, see also \cite{DKDH}. Finally, taking into
account that $S_{\infty}(x) = 0$ and $\sum_{n=1}^{\infty} W_{n}(x) = S_{1}(x)$,
we assure that the normalization condition holds true for an arbitrary $p(s)$.

\section{CHARACTERISTIC FUNCTION OF THE ARRIVAL TIME}

By use of the integral formula for the $\delta$ function,
\begin{equation}
    \delta(t - t_{x}^{(n)}) = \frac{1}{2\pi} \int_{-\infty}^{\infty}
    e^{-i\omega(t - t_{x}^{(n)})}d\omega,
    \label{int2}
\end{equation}
we can rewrite the probability density (\ref{Px(t)2}) in the form of a Fourier
integral, i.e.,
\begin{equation}
    P_{x}(t) = \frac{1}{2\pi}\int_{-\infty}^{\infty}\phi_{x}(\omega)
    \,e^{-i\omega t}d\omega.
    \label{Px(t)3}
\end{equation}
According to Eqs.~(\ref{Px(t)2})--(\ref{Px(t)3}), the characteristic function
$\phi_{x}(\omega)$, which determines all the statistical properties of the
arrival time $t_{x}$, is obtained as
\begin{equation}
    \phi_{x}(\omega) = e^{i\omega t_{x}^{(0)}}W_{0}(x) +
    \tilde{\phi}_{x}(\omega),
    \label{phi1}
\end{equation}
where
\begin{equation}
    \tilde{\phi}_{x}(\omega) = \sum_{n=1}^{\infty}
    \int_{\Omega_{n}(x)}e^{i\omega t_{x}^{(n)}} dW_{n}(x).
    \label{phi2}
\end{equation}

In the general case of an arbitrary $p(s)$, the characteristic function has a
complex structure involving an integration over the $n$-dimensional domain
$\Omega_{n}(x)$ and a summation over all $n$. Remarkably, however, $\phi_{x}
(\omega)$ can be expressed in terms of elementary functions if the random
intervals $s_{j}$ are exponentially distributed, i.e., if $p(s) = \lambda
e^{-\lambda s}$, where $\lambda^{-1}$ is the average length of $s_{j}$. In this
case the probability (\ref{dWn}) becomes
\begin{equation}
    dW_{n}(x) = e^{-\lambda x} \lambda^{n} \prod_{j = 1}^{n} ds_{j},
    \label{dWn2}
\end{equation}
and (\ref{phi2}) reduces to
\begin{equation}
    \tilde{\phi}_{x}(\omega) = e^{-\lambda x}\sum_{n=1}^{\infty}
    \lambda^{n}\int_{\Omega_{n}(x)}e^{i\omega t_{x}^{(n)}}
    \prod_{j=1}^{n}ds_{j}.
    \label{phi3}
\end{equation}

For the calculation of $\tilde{\phi}_{x}(\omega)$ it is convenient to transform
the right-hand side of Eq.~(\ref{phi3}) into a form with separate integrations
over the variables $s_{j}$. To this end, we use an approach \cite{DKDH} based
on the integral representation of the step function
\begin{equation}
    \frac{1}{2\pi}\int_{-\infty}^{+\infty}\frac{e^{(i\kappa + \eta) y}}
    {i\kappa + \eta}\,d\kappa = \left\{ \begin{array}{ll} 1 \;\;
    \textrm{if} \; y > 0 \\ [6pt] 0 \;\;\textrm{if} \; y < 0
    \end{array}
    \right.
    \label{int1}
\end{equation}
which is valid for  arbitrary $\eta > 0$. Applying (\ref{int1}) to (\ref{phi3})
and setting $y = x - \sum_{j=1}^{n} s_{j}$, one obtains the desired result:
\begin{eqnarray}
    \tilde{\phi}_{x}(\omega) \!\!&=&\!\! \frac{e^{-\lambda x}}{2\pi}
    \sum_{n=1}^{\infty}\lambda^{n} \int_{-\infty}^{\infty}d\kappa
    \frac{e^{(i\kappa + \eta)x}}{i\kappa + \eta}\int_{0}^{\infty}
    \!\!\ldots\!\int_{0}^{\infty}\! e^{i\omega t_{x}^{(n)}}
    \nonumber\\[6pt]
    && \!\!\times e^{-(i\kappa + \eta)\sum_{j=1}^{n} s_{j}}
    \prod_{j=1}^{n}ds_{j}.
    \label{phi4}
\end{eqnarray}

Then, using the identity $\sum_{n=1}^{\infty}a_{n} = \sum_{m=1}^{\infty}
[a_{2m-1} + a_{2m}]$ and taking into account that, according to (\ref{rel1a}),
\begin{eqnarray}
    &\displaystyle t_{x}^{(2m-1)} = \frac{x}{f+g} + \frac{2g}{f^{2} -
    g^{2}} \sum_{j=1}^{m}s_{2j-1},&
    \nonumber\\[3pt]
    &\displaystyle t_{x}^{(2m)} = \frac{x}{f-g} - \frac{2g}{f^{2} -
    g^{2}} \sum_{j=1}^{m}s_{2j},&
    \label{t^2m-1,t^2m}
\end{eqnarray}
we reduce the formula (\ref{phi4}) to the form
\begin{eqnarray}
    \tilde{\phi}_{x}(\omega) \!\!&=&\!\! \frac{e^{-\lambda x}}{2\pi}
    \int_{-\infty}^{\infty}\! \frac{e^{\nu_{0}x}}{\nu_{0}}\sum_{m=1}^
    {\infty} \Big[ I^{m-1}(\nu_{0})I^{m}(\nu_{1})e^{i\omega x/(f+g)}
    \nonumber\\[6pt]
    && \!\!+\, I^{m}(\nu_{0})I^{m}(\nu_{2})e^{i\omega x/(f-g)} \Big]
    d\kappa.
    \label{phi5}
\end{eqnarray}
Here,
\begin{equation}
    I(\nu_{k}) = \int_{0}^{\infty} p(s)e^{-(\nu_{k} - \lambda)s}ds =
    \frac{\lambda}{\nu_{k}}
    \label{def I}
\end{equation}
($\text{Re}\, \nu_{k} > 0$, $k = 0,1,2$) and
\begin{equation}
    \nu_{k} = i\kappa + \eta + i\frac{2q\omega}{f^{2} - g^{2}} \delta_{k}
    \label{nu k}
\end{equation}
with $\delta_{0} = 0$, $\delta_{1} = -1$, and $\delta_{2} = 1$. We note that
the right-hand side of (\ref{phi5}) contains an arbitrary positive parameter
$\eta$. According to the definition (\ref{phi2}), however, the left-hand side
of (\ref{phi5}) does not depend on $\eta$. This implies that the final result
of evaluating the series and integral in (\ref{phi5}) does not depend on $\eta$
as well. Therefore, for auxiliary manipulations we may choose a most convenient
value for this parameter.

From this point of view, it is reasonable to choose $\eta > \lambda$. This is
so because in this case $|I(\nu_{k})| < 1$ and the series in (\ref{phi5}) can
be easily evaluated:
\begin{eqnarray}
    &\displaystyle\sum_{m=1}^{\infty}I^{m-1}(\nu_{0})I^{m}(\nu_{1}) =
    \frac{\lambda \nu_{0}}{\nu_{0}\nu_{1} - \lambda^{2}},&
    \nonumber\\[6pt]
    &\displaystyle\sum_{m=1}^{\infty}I^{m}(\nu_{0})I^{m}(\nu_{2}) =
    \frac{\lambda^{2}}{\nu_{0}\nu_{2} - \lambda^{2}}.&
    \label{series}
\end{eqnarray}
Substituting (\ref{series}) into (\ref{phi5}) and using that $W_{0}(x) =
e^{-\lambda x}$ and
\begin{equation}
    e^{i\omega t_{x}^{(0)}}W_{0}(x) = \frac{e^{-\lambda x}}{2\pi}
    \int_{-\infty}^{\infty} \frac{e^{\nu_{0}x}}{\nu_{0}}\,
    e^{i\omega x/(f-g)}d\kappa,
    \label{rel2}
\end{equation}
from (\ref{phi1}) we obtain
\begin{equation}
    \phi_{x}(\omega) = \frac{e^{-\lambda x}}{2\pi}\int_{-\infty}^
    {\infty} e^{\nu_{0}x} \bigg[\frac{\lambda e^{i\omega x/(f+g)}}
    {\nu_{0}\nu_{1} - \lambda^{2}} + \frac{\nu_{2} e^{i\omega x/(f-g)}}
    {\nu_{0}\nu_{2} - \lambda^{2}} \bigg]d\kappa.
    \label{phi6}
\end{equation}

Upon calculating the integrals in (\ref{phi6}) (the details are given in the
Appendix) we find a remarkably simple expression for the characteristic
function of the arrival time
\begin{eqnarray}
    \phi_{x}(\omega) \!\!&=&\!\! e^{-\lambda x(1 - i\nu f/g)}
    \bigg[ \cosh\Big( \lambda x \sqrt{1 - \nu^2} \Big)
    \nonumber\\[6pt]
    && \!\!+\, \frac{1 + i\nu}{\sqrt{1 - \nu^2}} \sinh\Big(
    \lambda x \sqrt{1 - \nu^2} \Big) \bigg],
    \label{phi7}
\end{eqnarray}
where
\begin{equation}
    \nu = \frac{\omega g}{\lambda(f^2 - g^2)}.
    \label{nu}
\end{equation}
We note that $\phi_{x} (\omega)$, being the characteristic function, satisfies
the conditions $|\phi_{x}(\omega)| \leq 1$, $\phi_{x}(0) = 1$ and $\phi_{x}
(-\omega) = \phi_{x}^{*}(\omega)$, where the asterisk indicates the complex
conjugation. Equation (\ref{phi7}) is our main result which allows us to study
the essential properties of the arrival time analytically.

\section{MOMENTS OF THE ARRIVAL TIME}

The moments of the arrival time are defined in the usual way as $\langle
t_{x}^{m} \rangle = \int_{-\infty}^ {\infty} t^{m} P_{x}(t) dt$, and can be
deduced through the characteristic function as follows:
\begin{equation}
    \langle t_{x}^{m} \rangle = \frac{1}{i^{m}} \frac{d^{m}}
    {d\omega^{m}}\,\phi_{x}(\omega)\Big|_{\omega = 0}.
    \label{<t^m>}
\end{equation}
According to (\ref{phi7}) and (\ref{<t^m>}), the first moment, i.e.,
the mean arrival time, emerges as
\begin{equation}
    \langle t_{x} \rangle = \frac{1}{2\lambda(f^2-g^2)}\Big[2\lambda fx +
    g - ge^{-2\lambda x}\Big].
    \label{<t>}
\end{equation}
At small distances from the origin, when $\lambda x \ll 1$, the formula
(\ref{<t>}) yields $\langle t_{x} \rangle = x/(f-g)$. This result is  expected:
The total probability of those sample paths of $g(x)$ which do not change the
sign on the interval $(0,x)$ tends to 1 as $\lambda x \to 0$, and thus the
average particle velocity tends to $f-g$. In the other limiting case, when
$\lambda x \gg 1$, the formula (\ref{<t>}) approaches  $\langle t_{x} \rangle =
fx/(f^2-g^2)$. This result is corroborated by the fact that the long-time
asymptotic of the average particle velocity equals $(f^{2} - g^{2}) /f$
\cite{DKDH}.

The moments of higher order can also be calculated straightforwardly. In
particular, for the second moment we obtain
\begin{eqnarray}
    \langle t_{x}^{2} \rangle \!\!&=&\!\! \frac{1}{2\lambda^2(f^2-g^2)^2}
    \Big[ 2\lambda^2f^2x^2 + 2\lambda g(f+g)x - g^2
    \nonumber\\[6pt]
    && \!\!-\, g(2\lambda fx - g)e^{-2\lambda x} \Big].
    \label{<t^2>}
\end{eqnarray}

The central moments, $\langle (t_{x} - \langle t_{x} \rangle)^m \rangle$, can
be determined from the finite series
\begin{equation}
    \langle (t_{x} - \langle t_{x} \rangle)^m \rangle =
    \sum_{j=0}^{m}(-1)^{m-j}C_{m}^{j} \langle t_{x}^{l} \rangle
    \langle t_{x} \rangle^{m-l},
    \label{centr1}
\end{equation}
where $C_{m}^{j}$ is the binomial coefficient, or, alternatively, by the
formula
\begin{equation}
    \langle (t_{x} - \langle t_{x} \rangle)^m \rangle = \frac{1}
    {i^{m}}\frac{d^{m}} {d\omega^{m}}\,\phi_{x}(\omega)\,
    e^{-i\omega \langle t_{x} \rangle}\Big|_{\omega = 0}.
    \label{centr2}
\end{equation}
Specifically, using either of these definitions, the variance of the arrival
time, $\sigma_{x}^{2} = \langle (t_{x} - \langle t_{x} \rangle)^2 \rangle$, can
be written in the form
\begin{equation}
    \sigma_{x}^{2} = \frac{g^{2}}{4 \lambda^{2}(f^2 - g^2)^2}\Big[
    4\lambda x - 3 + 4e^{-2\lambda x} - e^{-4\lambda x} \Big].
    \label{sigma1}
\end{equation}
At short distances, when $\lambda x \ll 1$, the variance reduces to
\begin{equation}
    \sigma_{x}^{2} = \frac{4\lambda g^{2}}{3(f^2 - g^2)^2}x^3,
    \label{sigma2}
\end{equation}
and at long distances, when $\lambda x \gg 1$, it reduces to
\begin{equation}
    \sigma_{x}^{2} = \frac{g^{2}}{\lambda(f^2 - g^2)^2}x.
    \label{sigma3}
\end{equation}

Moreover, the central moments of the arrival time possess interesting scaling
properties. Namely, using (\ref{centr2}) with (\ref{phi7}) and (\ref{<t>}), one
obtains
\begin{equation}
    \langle (t_{x} - \langle t_{x} \rangle)^m \rangle = \bigg(\frac{g}
    {\lambda(f^2 - g^2)}\bigg)^{m} \Psi_{m}(\lambda x),
    \label{centr3}
\end{equation}
where
\begin{equation}
    \Psi_{m}(\lambda x) = e^{-\lambda x} \frac{d^{m}}{dz^{m}}\,
    \Phi(z,\lambda x)\Big|_{z = 0}
    \label{Psi}
\end{equation}
is a function of the single variable $\lambda x$, and
\begin{eqnarray}
    \Phi(z,\lambda x) \!\!&=&\!\! \exp \bigg(- z\, \frac{1-
    e^{-2\lambda x}}{2}\bigg) \bigg[ \cosh\Big( \lambda x
    \sqrt{1 + z^2} \Big)
    \nonumber\\[6pt]
    && \!\!+\, \frac{1 + z}{\sqrt{1 + z^2}} \sinh\Big(\lambda x
    \sqrt{1 + z^2} \Big) \bigg].
    \label{Phi}
\end{eqnarray}
Thus, the central moments exhibit a universal dependence on $f$, $g$ and
$\lambda$, i.e., $\langle (t_{x} - \langle t_{x} \rangle)^m \rangle \propto
[g/\lambda (f^2 - g^2)]^m$.

\section{PROPERTIES OF THE ARRIVAL TIME PROBABILITY DENSITY}

As it follows from (\ref{phi7}), the characteristic function tends to
$e^{-\lambda x + i\omega x /(f-g)}$ as $|\omega| \to \infty$. According to
(\ref{Px(t)3}), this suggests that the probability density of the arrival time
contains the $\delta$-singular contribution, i.e.,
\begin{equation}
    P_{x}(t) = \delta\left(t - \frac{x}{f-g}\right)e^{-\lambda x} +
    \tilde{P}_{x}(t),
    \label{Px(t)4}
\end{equation}
where
\begin{equation}
    \tilde{P}_{x}(t) = \frac{1}{2\pi} \int_{-\infty}^{\infty}
    \tilde{\phi}_{x}(\omega) e^{-i\omega t}d\omega
    \label{reg P1}
\end{equation}
denotes the regular part of $P_{x}(t)$ and
\begin{equation}
    \tilde{\phi}_{x}(\omega) = \phi_{x}(\omega) - e^{-\lambda x +
    i\omega x/(f-g)}.
    \label{tilde phi}
\end{equation}
Because the intensity of the $\delta$-singular part decreases exponentially
with increasing $x$, its contribution to $P_{x}(t)$ plays a crucial role only
at short distances from the origin. The regular part rules the behavior of
$P_{x}(t)$ at longer distances.

\subsection{Behavior at short distances}

At $\lambda x \ll 1$ we can obtain the probability density $P_{x}(t)$ in a
simple way, without the need to evaluate the integral in (\ref{reg P1}). To
this end, we first note that the $\delta$-singular part of $P_{x}(t)$ is formed
by those sample paths of $g(x)$ that do not change the sign on the interval
$(0,x)$. Accordingly, only the sample paths which have at least one change of
the sign on this interval do contribute to the regular part $\tilde{P}
_{x}(t)$. For small values of $\lambda x$ repeated changes of the sign are
unlikely. Therefore, in order to determine $\tilde{P} _{x}(t)$, we consider the
sample paths with a single change of the sign. In this case the probability
$\tilde{P}_{x}(t)dt$ is equal to $dW_{1} (x) = \lambda e^{-\lambda x} ds_{1}$
and, because $t = t_{x}^{(1)} = x/(f + g) + 2gs_{1} /(f^2 - g^2)$, the relation
$\tilde{P}_{x}(t)dt = dW_{1} (x)$ at $\lambda x \ll 1$ then yields
\begin{equation}
    \tilde{P}_{x}(t) = \frac{\lambda (f^2 - g^2)}{2g}.
    \label{reg P2}
\end{equation}
Finally, substituting (\ref{reg P2}) into (\ref{Px(t)4}), we find the
probability density of the arrival time at $\lambda x \ll 1$:
\begin{equation}
    P_{x}(t) = \delta\left(t - \frac{x}{f-g}\right)(1-\lambda x) +
    \frac{\lambda(f^2 - g^2)}{2g}.
    \label{Px(t)5}
\end{equation}

At first sight, this result may come as a surprise because the regular part of
the probability density does not depend explicitly on $x$ and $t$. It should be
stressed, however, that the formula (\ref{reg P2}) is derived under the
condition that $t \in [t_{\text{min}}, t_{\text{max}}]$, where $t_{\text{min}}
= x/(f + g)$ and $t_{\text{max}} = x/(f - g)$ [we recall that $\tilde{P}_{x}(t)
\equiv 0$ if $t \notin [t_{\text{min}}, t_{\text{max}}]$]. This means that
$\tilde{ P}_{x}(t)$ depends on $x$ and $t$ implicitly leading to the broadening
of $\tilde{ P}_{x} (t)$ if $x$ increases. At the starting point $x = 0$ we have
$t_{\text{min}} = t_{\text{max}} = 0$, therefore $\tilde{ P}_{0} (t) = 0$ and,
in accordance with the condition $t_{0}= 0$, $P_{0}(t) = \delta(t)$. We note
also that the normalization condition $\int_{t_{ \text{ min}}}^ {t_{\text
{max}}} P_{x}(t) dt = 1$, which holds true also for (\ref{Px(t)5}), further
corroborates the validity of (\ref{reg P2}).

\subsection{Behavior at long distances}

To study the long-distance behavior of the probability density $P_{x}(t)$, it
is convenient to introduce the new time variable $\tau = (t - \langle t_{x}
\rangle)/ \sigma_{x}$. The corresponding scaled probability density $\mathcal
{P}_{x}(\tau)$ is expressed through $P_{x}(t)$ as $\mathcal {P}_{x} (\tau) =
\sigma_{x} P_{x}(\langle t_{x} \rangle + \sigma_{x} \tau)$ and, according to
(\ref{Px(t)3}), it can be written in the form
\begin{equation}
    \mathcal{P}_{x}(\tau) = \frac{1}{2\pi} \int_{-\infty}^{\infty} \phi_{x}
    (\mu/\sigma_{x})\,e^{-i\mu \langle t_{x} \rangle /\sigma_{x} -
    i\mu \tau} d\mu.
    \label{reducedP1}
\end{equation}
Using (\ref{phi7}), (\ref{<t>}) and (\ref{sigma1}), the characteristic function
of $\mathcal{P}_{x}(\tau)$ then reads
\begin{equation}
    \phi_{x}(\mu/\sigma_{x})\,e^{-i\mu \langle t_{x} \rangle /\sigma_{x}} =
    e^{-\lambda x}\Phi(i\mu\Psi_{2}^{-1/2}(\lambda x),\lambda x),
    \label{char red}
\end{equation}
where the function $\Phi(z,\lambda x)$ is defined by Eq.~(\ref{Phi}) and
\begin{equation}
    \Psi_{2}(\lambda x) = \lambda x - \frac{3}{4} + e^{-2\lambda x}
    -\frac{1}{4}\,e^{-4\lambda x}.
    \label{Psi2}
\end{equation}
Because the characteristic function (\ref{char red}) depends only on $\lambda
x$ and the integration variable $\mu$, the scaled probability density
(\ref{reducedP1}) possesses the remarkable property that $\mathcal {P}_{x}
(\tau)$ is a function of $\lambda x$ and $\tau$ which depends neither on the
external force $f$ nor on the amplitude $g$ of the dichotomous random force
$g(x)$.

In the case of long distances, if  $\lambda x \gg 1$, the characteristic
function (\ref{char red}) at $\mu^4 \ll \lambda x$ can be approximated by the
two terms of its expansion:
\begin{equation}
    \phi_{x}(\mu/\sigma_{x})\,e^{-i\mu \langle t_{x} \rangle /\sigma_{x}}
    = e^{-\mu^2/2}\left(1 - \frac{\mu^4}{8\lambda x} \right).
    \label{rel4}
\end{equation}
Substituting (\ref{rel4}) into (\ref{char red}) and calculating the integrals,
we find the two terms of expression of the scaled probability density
\begin{equation}
    \mathcal{P}_{x}(\tau) = \frac{e^{-\tau^2/2}}{\sqrt{2\pi}}
    \left(1 - \frac{3 - 6\tau^2 + \tau^4}{8\lambda x} \right)
    \label{reducedP2}
\end{equation}
which is valid if $\lambda x \gg \max{(1,\tau^4)}$. Thus, in accordance with
the central limit theorem of probability theory (see, e.g., Ref. \cite{GK}),
the limiting probability density approaches a Gaussian form, i.e., $\mathcal{P}
_{\infty} (\tau) = (2\pi)^{-1/2} e^{ -\tau^2/2}$, and $\mathcal{P}_{x} (\tau) -
\mathcal {P}_{\infty} (\tau) \propto (\lambda x)^{-1}$ as $\lambda x \to
\infty$.

\subsection{Numerical verification}

Our numerical calculations pursue two goals, namely (i) to verify the
analytical findings and  (ii) to  illustrate and visualize the obtained
findings. The former is achieved by comparison of the probability density
(\ref{Px(t)4}) with that derived from the numerical simulation of the arrival
time (\ref{arr time}). We use the Maple package for calculating the Fourier
integral in (\ref{reg P1}) and employ the histogram procedure to numerically
evaluate the probability density. In short, this procedure consists of
successive generations of random intervals $s_{j}$ according to the exponential
distribution and evaluating the arrival time to a fixed position $x$ for
different realizations of random intervals. The probability density is then
presented as the histogram of arrival times of the particle. For further
details about this procedure we refer the interested  reader to
Ref.~\cite{DKDH} where a similar approach was used for the numerical evaluation
of the probability density of the particle position at a fixed  time $t$. In
doing so, we made sure that the simulated probability density function is in
perfect agreement with the theoretical one, see Fig.~2.

Figure 3 illustrates the short-distance behavior of the probability density
(\ref{Px(t)4}). As can be seen  in Fig.~3(a), at very small values of $x$ the
probability density of the arrival time is described by the approximate formula
(\ref{Px(t)5}). In accordance with the assumption made in its derivation, this
suggests that the sample paths of $g(x)$ which on the interval $(0,x)$ have
more than one change of the sign are responsible for the explicit dependence of
$\tilde {P}_{x}(t)$ on $t$. If $x$ is not too small, i.e., the total
probability of these sample paths is small but non-zero, then, as shown in
Fig.~3(b), $\tilde {P}_{x}(t)$ is an almost linear function of $t$. With
increasing of $x$ the role of these sample paths becomes increasingly
important: The function $\tilde {P}_{x}(t)$ becomes nonlinear, assumes a
unimodal form, and eventually approaches a Gaussian shape, see Fig.~4.

\subsection{Skewness and kurtosis}

In order to quantitatively describe the difference between the arrival time
probability density and a Gaussian density with identical  mean and variance as
$P_{x} (t)$, we calculate the skewness
\begin{equation}
    s(x) = \frac{\langle(t_{x} - \langle t_{x} \rangle)^{3}\rangle}
    {\sigma_{x}^{3}}
    \label{s(x)1}
\end{equation}
that characterizes the degree of asymmetry of $P_{x} (t)$, and as well the
kurtosis
\begin{equation}
    k(x) = \frac{\langle(t_{x} - \langle t_{x} \rangle)^{4}\rangle}
    {\sigma_{x}^{4}} - 3
    \label{k(x)1}
\end{equation}
that characterizes the degree of peakedness of $P_{x} (t)$. Because $s(x)
\equiv 0$ and $k(x) \equiv 0$ if the arrival time $t_{x}$ follows a Gaussian
distribution, one can consider the skewness and kurtosis as appropriate
measures of deviation of the arrival time distribution from a Gaussian shape.
Using the representation (\ref{centr3}) for the central moments, from the
definitions (\ref{s(x)1}) and (\ref{k(x)1}) we obtain
\begin{equation}
    s(x) = \frac{\Psi_{3}(\lambda x)}{\Psi_{2}^{3/2} (\lambda x)},
    \quad
    k(x) = \frac{\Psi_{4}(\lambda x)}{\Psi_{2}^{2} (\lambda x)} - 3,
    \label{ks}
\end{equation}
i.e., $s(x)$ and $k(x)$ are universal functions of the single
variable $\lambda x$, see Fig.~5. Calculating $\Psi_{3} (\lambda x)$
and $\Psi_{4} (\lambda x)$ and taking into account (\ref{Psi2}), we
find an explicit expression for the skewness,
\begin{eqnarray}
    s(x) \!\!&=&\!\! -\frac{2}{(4\lambda x - 3 + 4e^{-2\lambda x} -
    e^{-4\lambda x})^{3/2}}\,[2 + (3
    \nonumber\\[6pt]
    && \!\! -\, 12\lambda x)e^{-2\lambda x} - 6e^{-4\lambda x} +
    e^{-6\lambda x}],
    \label{s(x)2}
\end{eqnarray}
and for the kurtosis,
\begin{eqnarray}
    k(x) \!\!&=&\!\! \frac{6}{(4\lambda x - 3 + 4e^{-2\lambda x} -
    e^{-4\lambda x})^{2}}\,[13 - 8\lambda x
    \nonumber\\[6pt]
    && \!\! -\, 8(1+4\lambda x)e^{-2\lambda x} - 4(3 - 4\lambda x)
    e^{-4\lambda x}
    \nonumber\\[6pt]
    && \!\! +\, 8e^{-6\lambda x} - e^{-8\lambda x}].
    \label{k(x)2}
\end{eqnarray}

The formulas (\ref{s(x)2}) and (\ref{k(x)2}) yield in leading order of $\lambda
x$ the following relations: $s(x) = - (3\sqrt{3}/ 4) (\lambda x)^{-1/2}$ and
$k(x) = (9/5)(\lambda x)^{-1}$ at $\lambda x \ll 1$, and $s(x) = -(1/2)
(\lambda x)^{-3/2}$ and $k(x) = -3(\lambda x)^{-1}$ at $\lambda x \gg 1$. These
results clearly evidence that the arrival time probability density $P_{x} (t)$
distinctly differs from a Gaussian density at short distances and approaches
this Gaussian shape at long distances. Moreover, since $|k(x)/s(x)| \to \infty$
as both, $\lambda x \to 0$ and $\lambda x \to \infty$, the kurtosis can be
considered as a unique measure of non-Gaussianity of $P_{x}(t)$. We note that,
because of the condition $s(x) < 0$, the left tail of $P_{x}(t)$ is always
heavier than the right tail. Also, $P_{x}(t)$ is more peaked compared to the
Gaussian density at distances where $k(x) > 0$, and is more flattened at
distances where $k(x) <0$.

\section{CONCLUSIONS}

We applied the path integral approach to calculate the characteristic function
of the arrival time for overdamped particles driven by a constant bias in a
piecewise linear random potential producing a dichotomous random force with
exponentially distributed spatial intervals. Using the characteristic function,
we derived the moments of the arrival time, established  universal scaling
properties of the central moments, and demonstrated that the arrival time
probability density $P_{x}(t)$ contains both a $\delta$-singular contribution
and a regular part. While the $\delta$-singular part, whose weight decreases
exponentially with increasing $x$, plays the main role at short distances, the
regular part of $P_{x}(t)$ dominates at large distances $x$.

At very small distances the regular part is defined by the sample paths with
only one change of the sign on the interval $(0,x)$ and in this case its value
does not depend on $x$. Upon increasing $x$ the contribution of other sample
paths leads to the transformation of this part of $P_{x}(t)$ into an almost
linear function of $t$, and subsequently  into unimodal form, and finally, at
$x \to \infty$, it tends to a Gaussian density as $x^{-1}$. Moreover, in order
to characterize the difference of the arrival time probability density from the
Gaussian density, we calculated the skewness and kurtosis. The function
$P_{x}(t)$ is more peaked in comparison with the Gaussian density at small $x$
and is more flattened at large $x$.

\section*{ACKNOWLEDGMENTS}

S.I.D. acknowledges the support of the EU through Contract No
MIF1-CT-2006-021533 and P.H. acknowledges the support by the Deutsche
For\-schungs\-ge\-mein\-schaft via the Collaborative Research Centre SFB-486,
project A10.  Financial support from the German Excellence Initiative via the
{\it Nanosystems Initiative Munich} (NIM) is gratefully acknowledged as well.

\appendix*

\section{DERIVATION OF THE CHARACTERISTIC FUNCTION}

For calculating the integrals in (\ref{phi6}),
\begin{eqnarray}
    &\displaystyle Y = \frac{1}{2\pi}\int_{-\infty}^{\infty}\frac{\lambda
    e^{\nu_{0}x}}{\nu_{0}\nu_{1} - \lambda^{2}} d\kappa,&
    \nonumber\\[6pt]
    &\displaystyle Z = \frac{1}{2\pi}\int_{-\infty}^{\infty}\frac{\nu_{2}
    e^{\nu_{0}x}}{\nu_{0}\nu_{2} - \lambda^{2}} d\kappa,&
    \label{YZ1}
\end{eqnarray}
we use the method of contour integration \cite{MF}. According to (\ref{nu k})
and (\ref{nu}), the integrands in (\ref{YZ1}), $R(\kappa) = \lambda
e^{\nu_{0}x} /(\nu_{0} \nu_{1} - \lambda^{2})$ and $S(\kappa) =
\nu_{2}e^{\nu_{0}x} /(\nu_{0}\nu_{2} - \lambda^{2})$, can be written in the
form
\begin{eqnarray}
    &\displaystyle R(\kappa) = - \frac{\lambda}{(\kappa - \kappa_{1})
    (\kappa -\kappa_{2})}\,e^{(i\kappa + \eta)x},&
    \nonumber\\[6pt]
    &\displaystyle S(\kappa) = - \frac{i\kappa + \eta + 2i\lambda
    \nu}{(\kappa - \kappa_{3})(\kappa -\kappa_{4})}\,
    e^{(i\kappa + \eta)x},&
    \label{RS}
\end{eqnarray}
where
\begin{eqnarray}
    &\kappa_{1,2} = i\eta \pm i\lambda \sqrt{1 - \nu^2} + \lambda \nu,&
    \nonumber\\[6pt]
    &\kappa_{3,4} = i\eta \pm i\lambda \sqrt{1 - \nu^2} - \lambda \nu.&
    \label{kappa1-4}
\end{eqnarray}
The formulas (\ref{RS}) exhibit that both $R(\kappa)$ and $S(\kappa)$ as
functions of the complex variable $\kappa$ have two poles of the first order at
$\kappa = \kappa_{1,2}$ and $\kappa = \kappa_{3,4}$, respectively. If $\eta >
\lambda$ then all poles are located in the upper half plane of the complex
$\kappa$ plane, and the residue theorem yields
\begin{eqnarray}
    &Y = i[\text{Res}\,R(\kappa_{1}) + \text{Res}\, R(\kappa_{2})],&
    \nonumber\\[6pt]
    &Z = i[\text{Res}\,S(\kappa_{3}) + \text{Res}\, S(\kappa_{4})].&
    \label{YZ2}
\end{eqnarray}
Since the residues in (\ref{YZ2}) are defined as $\text{Res}\,R(\kappa_{1,2}) =
\lim_{\kappa \to \kappa_{1,2}} (\kappa - \kappa_{1,2})R(\kappa)$ and
$\text{Res}\,S(\kappa_{3,4}) = \lim_{\kappa \to \kappa_{3,4}} (\kappa -
\kappa_{3,4})S(\kappa)$, from (\ref{RS}) and (\ref{kappa1-4}) we obtain
\begin{eqnarray}
    Y \!\!&=&\!\! \frac{1}{\sqrt{1 - \nu^2}}\sinh\Big( \lambda x
    \sqrt{1 - \nu^2} \Big) e^{i\lambda x\nu},
    \nonumber\\[8pt]
    Z \!\!&=&\!\!
    \bigg[ \frac{i\nu}{\sqrt{1 - \nu^2}} \sinh\Big( \lambda x
    \sqrt{1 - \nu^2} \Big)
    \nonumber\\[6pt]
    && \!\!+\, \cosh\Big( \lambda x \sqrt{1 - \nu^2}\Big)\bigg]
    e^{-i\lambda x\nu}.
    \label{YZ3}
\end{eqnarray}
Finally, substituting (\ref{YZ3}) into the formula
\begin{equation}
    \phi_{x}(\omega) = e^{-\lambda x} \Big[ Ye^{i\omega x/(f+g)} +
    Ze^{i\omega x/(f-g)} \Big]
    \label{phi}
\end{equation}
which follows from (\ref{phi6}) and (\ref{YZ1}), we get the desired
characteristic function (\ref{phi7}).

\newpage

\begin{figure}
    \centering
    \includegraphics[totalheight=6cm]{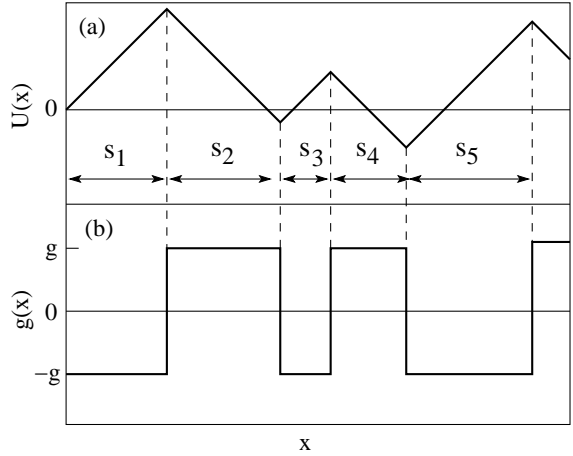}
    \caption{\label{fig1} Schematic representation of (a) the piecewise
    linear random potential $U(x)$ and (b) the corresponding dichotomous
    random force $g(x) = -dU(x)/ dx$ as functions of the coordinate $x$. }
\end{figure}

\begin{figure}
    \centering
    \includegraphics[totalheight=4.5cm]{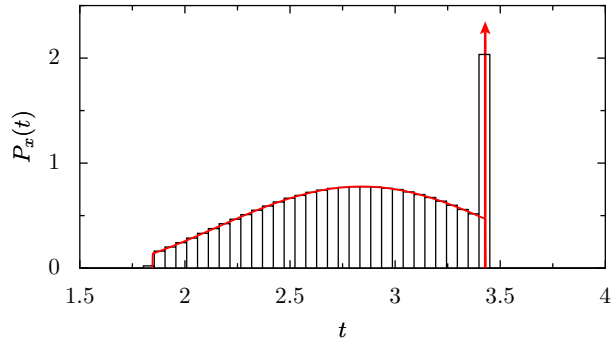}
    \caption{\label{fig2} (Color online) Theoretical and simulated
    probability density of the arrival time for $x = 2.4$. The solid
    line (red online) and histogram represent the analytical result
    (\ref{Px(t)4}) and the numerical simulation of the arrival
    time (\ref{arr time}), respectively. The parameters of the
    force field are chosen as $f = 1$, $g = 0.3$, and $\lambda = 1$.
    The vertical arrow depicts the $\delta$-singular contribution
    to $P_{x}(t)$. }
\end{figure}

\begin{figure}
    \centering
    \includegraphics[totalheight=4.5cm]{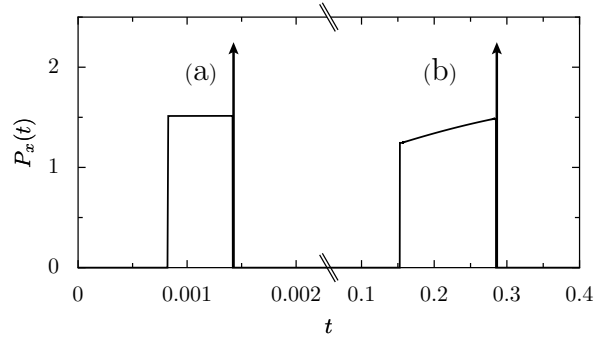}
    \caption{\label{fig3} Short-distance behavior of the probability
    density of the arrival time $P_x(t)$ at (a) $x = 10^{-3}$ and (b)
    $x = 0.2$. The other parameters are the same as those in Fig.~2. }
\end{figure}

\begin{figure}
    \centering
    \includegraphics[totalheight=4.5cm]{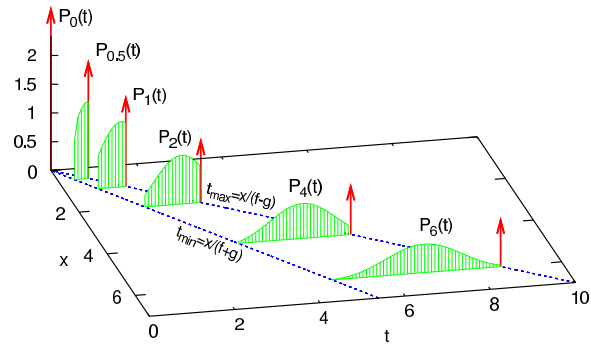}
    \caption{\label{fig4} (Color online) Plots of the probability density
    of the arrival time for different distances from the origin of the
    coordinate system. The vertical surfaces (green online) depict the
    regular part of $P_{x}(t)$. In order to visually demonstrate that the
    intensity of the $\delta$-singular part of $P_{x}(t)$ exponentially
    decreases with $x$, we depicted the length of the vertical arrows
    (red online) in the form $0.8 + 2e^{-\lambda x}$. For the convenience
    of comparison, the force field characteristics are chosen as in
    Figs.~2 and 3. }
\end{figure}

\begin{figure}
    \centering
    \includegraphics[totalheight=4.5cm]{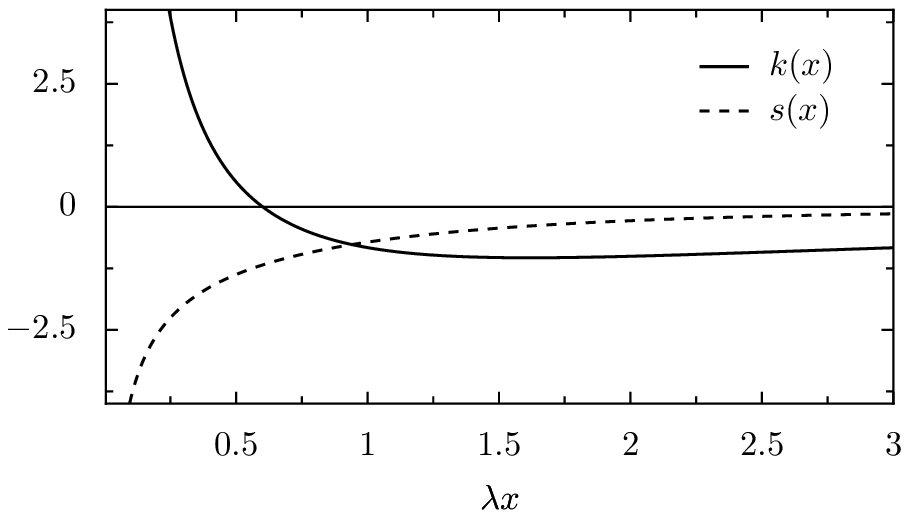}
    \caption{\label{fig5} Plots of the skewness $s(x)$ (dashed line)
    and the kurtosis $k(x)$ (solid line) of the arrival time probability
    density vs the normalized distance $\lambda x$. }
\end{figure}


\begin{thebibliography}{99}

\bibitem{HL}
W.~Horsthemke and R.~Lefever, \textit{Noise-Induced Transitions}
(Springer-Verlag, Berlin, 1984).

\bibitem{Trans}
P.~Reimann and P. H\"{a}nggi, Appl.\ Phys.\ A:\ Mater.\ Sci.\ Process.\
\textbf{75}, 169 (2002); R.~D.~Astumian and P.~H\"{a}nggi, Phys.\ Today {\bf
55} (11), 33 (2002); P.~H\"{a}nggi, F.~Marchesoni, and F.~Nori, Ann.\ Phys.\
(Leipzig) {\bf 14}, 51 (2005).

\bibitem{GHJM}
L.~Gammaitoni, P.~H\"{a}nggi, P.~Jung, and F.~Marchesoni, Rev.\ Mod.\ Phys.\
\textbf{70}, 223 (1998).

\bibitem{BG}
J.-P.~Bouchaud and A.~Georges, Phys.\ Rep.\ \textbf{195}, 127 (1990).

\bibitem{Sin}
Ya.~G.~Sinai, Theor.\ Probab.\ Appl.\ \textbf{27}, 256 (1982).
\bibitem{Der}
B.~Derrida, J.\ Stat.\ Phys.\ \textbf{31}, 433 (1983).
\bibitem{Gol}
A.~O.~Golosov, Commun.\ Math.\ Phys.\ \textbf{92}, 491 (1984).
\bibitem{Mon}
C.~Monthus, Lett.\ Math.\ Phys.\ \textbf{78}, 207 (2006).

\bibitem{HTB}
P.~H\"{a}nggi, P.~Talkner, and M.~Borkovec, Rev.\ Mod.\ Phys.\ \textbf{62}, 251
(1990), see Sec. VII, C.3.
\bibitem{Sch}
S.~Scheidl, Z.\ Phys.\ B\ \textbf{97}, 345 (1995).
\bibitem{DV}
P.~Le Doussal and V.~M.~Vinokur, Physica\ C\ \textbf{254}, 63 (1995).
\bibitem{GB}
D.~A.~Gorokhov and G.~Blatter, Phys.\ Rev.\ B\ \textbf{58}, 213 (1998).
\bibitem{LV}
A.~V.~Lopatin and V.~M.~Vinokur, Phys.\ Rev.\ Lett.\ \textbf{86}, 1817 (2001).

\bibitem{Den}
S.~I.~Denisov, J.\ Magn.\ Magn.\ Mater.\ \textbf{147}, 406 (1995).
\bibitem{DL}
S.~I.~Denisov and R.~Yu.~Lopatkin, Phys.\ Scr.\ \textbf{56}, 423 (1997).
\bibitem{PKDK}
P.~E.~Parris, M.~Ku\'{s}, D.~H.~Dunlap, and V.~M.~Kenkre, Phys.\ Rev.\ E\
\textbf{56}, 5295 (1997).
\bibitem{KKDP}
V.~M.~Kenkre, M.~Ku\'{s}, D.~H.~Dunlap, and P.~E.~Parris, Phys.\ Rev.\ E\
\textbf{58}, 99 (1998).
\bibitem{DH}
S.~I.~Denisov and W.~Horsthemke, Phys.\ Rev.\ E\ \textbf{62}, 3311 (2000).

\bibitem{PASF}
M.~N.~Popescu, C.~M.~Arizmendi, A.~L.~Salas-Brito, and F.~Family, Phys.\ Rev.\
Lett.\ \textbf{85}, 3321 (2000).
\bibitem{GLZH}
L.~Gao, X.~Luo, S.~Zhu, and B.~Hu, Phys.\ Rev.\ E\ \textbf{67}, 062104 (2003).
\bibitem{ZLAF}
D.~G.~Zarlenga, H.~A.~Larrondo, C.~M.~Arizmendi, and F. Family, Phys.\ Rev.\ E\
\textbf{75}, 051101 (2007).

\bibitem{DKDH}
S.~I.~Denisov, M.~Kostur, E.~S.~Denisova, and P. H\"{a}nggi, Phys.\ Rev.\ E\
\textbf{75}, 061123 (2007).

\bibitem{GK}
B.~V.~Gnedenko and A.~N.~Kolmogorov, \textit{Limit Distributions for Sums of
Independent Random Variables} (Addison-Wesley, Cambridge, MA, 1954).

\bibitem{MF}
P.~M.~Morse and H.~Feshbach, \textit{Methods of Theoretical Physics}
(McGraw-Hill, New York, 1953), Vol.~1.

\end{thebibliography}
\end{document}